\documentclass[journal]{IEEEtran}

\usepackage{amssymb,amsmath}
\usepackage{graphicx,tabularx,multirow}
\usepackage{cite,url}
\usepackage{color}
\usepackage{algorithm}
\usepackage{algorithmic}
\usepackage{subcaption,epstopdf}
\usepackage{tikz}
\usepackage{cancel}
\usepackage[percent]{overpic}

\newenvironment{remark}{\par\medskip
   \noindent \textit{Remark.} \rmfamily}{\medskip}

\newcounter{claim}
\newenvironment{claim}{\refstepcounter{claim}\par\medskip
   \noindent  \textbf{Claim \theclaim.} \em \rmfamily}
{\medskip}
   
\newenvironment{example}[1][]{%\par\medskip
   \noindent \textit{Example (#1).} \rmfamily}{\hfill $\square$}%{\medskip}

\newenvironment{myproof}{%\par\medskip
   \noindent \textit{Proof.} \rmfamily}{\hfill $\square$\medskip}

\newenvironment{myproof*}{%\par\medskip
   \noindent \textit{Proof.} \rmfamily}{\hfill $\blacksquare$}

\newcounter{theorem}

\newcounter{corollary}
\newenvironment{corollary}{\refstepcounter{corollary}\par\medskip
   \noindent  \textbf{Corollary \thecorollary.} \em \rmfamily}{\medskip}

\newcounter{lemma}

\newcounter{proposition}
\newenvironment{proposition}{\refstepcounter{proposition}\par\medskip
   \noindent  \textbf{Proposition \theproposition.} \em \rmfamily}
{\medskip}

\DeclareMathOperator*{\argmin}{argmin\;}

\newcommand{\nn}{\nonumber}
\newcommand{\universe}[1]{\mathbb{#1}}
\newcommand{\set}[1]{\mathcal{#1}}
\newcommand{\player}[1]{$\mathcal{P}_{#1}$}
\newcommand{\cost}[1]{c_{#1}}

\newcommand{\trace}[1]{\mathrm{tr}\left\{#1\right\}}
\newcommand{\rand}[1]{\pmb{#1}}

\newcommand{\rank}[1]{\mathrm{rank}\left\{#1\right\}}

\newcommand{\co}[1]{\mathrm{co}\left\{#1\right\}}

\newcommand{\estimate}[1]{\hat{\rand{#1}}}
\renewcommand{\vec}[1]{\underline{#1}}
\newcommand{\E}[1]{\universe{E}\left\{#1\right\}}
\newcommand{\prob}[1]{\mathbb{P}\left\{#1\right\}}

\newcommand{\diag}[1]{\mathrm{diag}\left\{#1\right\}}

\newcommand{\interior}[1]{\mathrm{int}\left\{#1\right\}}

\newcommand{\subjectTo}{~\mathrm{s.t.}}
\newcommand{\CP}{\mathcal{CP}}
\newcommand{\COP}{\mathcal{COP}}

\newcommand{\state}{z}
\newcommand{\info}{x}
\newcommand{\private}{y}
\newcommand{\signal}{s}
\newcommand{\signalsSet}{\set{S}}

\newcommand{\infoSet}{\set{X}}
\newcommand{\stateSet}{\set{Z}}
\newcommand{\privateSet}{\set{Y}}

\newcommand{\Srule}{\pi}
\newcommand{\Sspace}{\Pi}
\newcommand{\Rrule}{\gamma}
\newcommand{\Rspace}{\Gamma}

\newcommand{\prior}{p_o}
\newcommand{\posterior}{p_{\signal}}
\newcommand{\myXi}{\Xi_{\Srule}}
\newcommand{\myPi}{\underline{\Pi}}

\newenvironment{psmallmatrix}
  {\left[\begin{smallmatrix}}
  {\end{smallmatrix}\right]}

\newcommand{\cb}{}

\begin{document}

%\title{Deceptive Signaling for General Class of Distributions}
%\title{A Copositive Program Solution of Bayesian Persuasion for Quadratic Cost Measures}
%\title{Optimal Hierarchical Signaling for\\ Quadratic Cost Measures and General Distributions:\\
%A Copositive Program Characterization}
\title{\cb Bayesian Persuasion with State-Dependent Quadratic Cost Measures}

\author{Muhammed~O.~Sayin and~Tamer~Ba\c{s}ar,~\IEEEmembership{Life~Fellow,~IEEE}
\thanks{This research was supported in part by the U.S. Office of Naval Research (ONR) MURI grant N00014-16-1-2710, in part by the U.S. Army Research Office (ARO) MURI grant AG285, and in part by the U.S. Army Research Labs (ARL) under IoBT Grant 479432-239012-191100.

Muhammed O. Sayin is with the Laboratory for Information and Decision Systems, Massachusetts Institute of Technology, Cambridge, MA 02139. E-mail: sayin@mit.edu

Tamer Ba\c{s}ar is with the Department of Electrical and Computer Engineering, University of Illinois at Urbana-Champaign, Urbana, IL 61801 USA. E-mail: basar1@illinois.edu}}

\maketitle

\begin{abstract}\cb
We address Bayesian persuasion between a sender and a receiver with state-dependent quadratic cost measures for general classes of distributions. The receiver seeks to make mean-square-error estimate of a state based on a signal sent by the sender while the sender signals strategically in order to control the receiver's estimate in a certain way. Such a scheme could model, e.g., deception and privacy, problems in multi-agent systems. Existing solution concepts are not viable since here the receiver has continuous action space. We show that for finite state spaces, optimal signaling strategies can be computed through an equivalent linear optimization problem over the cone of completely positive matrices. We then establish its strong duality to a copositive program. To exemplify the effectiveness of this equivalence result, we adopt sequential polyhedral approximation of completely-positive cones and analyze its performance numerically. We also quantify the approximation error for a quantized version of a continuous distribution and show that a semi-definite program relaxation of the equivalent problem could be a benchmark lower bound for the sender's cost for large state spaces.
\end{abstract}


\section{Introduction}

\IEEEPARstart{T}{o} induce intelligent decision makers to take certain actions, we can create incentives via external means, e.g., explicit payments, but we can also persuade them to take the actions by their own will without the need for any external means if we can craft the information available to them. {\cb In the field of economics,} such a problem of interest has been examined by Kamenica and Gentzkow in their seminal paper \cite{ref:Kamenica11}. {\cb In particular, they have introduced the ``Bayesian persuasion" problem between a sender and a receiver, where the sender crafts the signal sent to the receiver in order to induce the receiver to take a certain action while the receiver is aware of how the signal is crafted (e.g., due to committed transparency/honesty).} They have provided a geometrical interpretation of the optimal persuasive signaling strategy and examined the persuasion capability of the sender. A detailed justification of the sender's policy commitment can be found in \cite{ref:Kamenica11} and a recent survey of literature on Bayesian persuasion could be found in \cite{ref:Kamenica19}. 

{\cb Applications that can be modeled within Bayesian persuasion framework are not limited to the field of economics, since information sharing is a common theme also in multi-agent communication (and control) systems. Further, in non-cooperative multi-agent systems, (selfish) agents with diverse objectives are likely to have tendency to share information strategically. Correspondingly, strategic signaling has attracted significant attention also in the fields of {\em communication} and {\em control}, recently, with a specific focus on {\em state-dependent quadratic cost} measures (due to their wide applicability in communication and control systems).} However, these studies have mainly focused on Gaussian information models \cite{ref:Farokhi17, ref:Tamura14,ref:Akyol17,ref:Saritas17, ref:Sayin17b, ref:Sayin19a, ref:Sayin19c} (different from the literature in economics \cite{ref:Kamenica11,ref:Gentzkow16,ref:Kolotilin18,ref:Dworczak19} and computer science \cite{ref:Dughmi16,ref:Dughmi17}).

In \cite{ref:Farokhi17}, the authors have shown that a certain linear signaling strategy is optimal provided that sender commits to a signaling policy beforehand and receiver has bounded rationality by using linear estimates only. Indeed, for the same settings of \cite{ref:Farokhi17}, in \cite{ref:Tamura14}, the author had shown that the linear signaling strategy is optimal even when the receiver is completely rational by selecting any measurable decision policy. {\cb We note that without the policy commitment of the sender, there does not exist a Nash equilibrium where the sender selects a linear signaling strategy. For example, an early reference \cite{ref:Crawford82} had shown\footnote{\cb In \cite{ref:Crawford82}, the authors have focused only on compact state spaces and quite general class of cost measures including, but not limited to, quadratic costs. Later, \cite{ref:Saritas17} has shown that the result can be generalized to settings where the state is drawn according to a Gaussian distribution (which does not have a compact support) and players have quadratic cost measures.} that there only exists partition (Nash) equilibrium (where sender sends a quantized version of the state) if there is no policy commitment.} Under the same settings with \cite{ref:Farokhi17} yet for a completely rational receiver, \cite{ref:Akyol17} has shown the optimality of linear signaling strategies when there is an additive Gaussian noise channel and the sender has a hard power constraint. {\cb Contrary to the noiseless channel scenarios, however, when there is no policy commitment, in \cite{ref:Saritas17} the authors have shown that under certain conditions the sender can also have a linear signaling strategy at a (Nash) equilibrium if there is an additive Gaussian noise channel and the sender has a soft power constraint on the signal sent.} 
 
In our previous papers \cite{ref:Sayin17b,ref:Sayin19a,ref:Sayin19c}, we have studied the strategic signaling problem described above in {\em dynamic} environments over a finite horizon. For discrete-time (multivariate) Gauss Markov processes, in \cite{ref:Sayin17b} we have shown the optimality of linear signaling strategies within the general class of measurable policies and formulated an equivalent semi-definite program (SDP), which enables computation of the optimal signaling strategies numerically through existing SDP solvers. For non-cooperative linear quadratic Gaussian (LQG) control problems, in \cite{ref:Sayin19a}, we have formulated optimal linear signaling strategies of a sensor who seeks to deceive a private-type controller in settings where the distribution over the types of the controller is not known. And reference \cite{ref:Sayin19c} provides an overview of the results of\cite{ref:Sayin17b} and \cite{ref:Sayin19a} {\cb from the viewpoint of security in cyber-physical systems.}

For Gaussian information, we can obtain well structured results, e.g., optimality of linear signaling strategies, also in dynamic and noisy environments, as shown in the studies reviewed above. However, for distributions other than Gaussian, we still have significant but not completely explored problems. Notably, for general distributions with compact support, \cite{ref:Kamenica11} brings in a geometrical interpretation into the problem, which requires the computation of a convex envelope of a function, which can be prominently challenging even for finite yet relatively large state spaces \cite{ref:Tardella04}. We also note that, as studied in \cite{ref:Gentzkow16,ref:Kolotilin18,ref:Dworczak19}, a relatively simpler characterization of the solution is possible for a special class of Bayesian persuasion problem{\cb, e.g., where the sender's cost measure does {\em not} depend on the state.} 

\begin{figure}[t!]
\centering
\includegraphics[width=.4\textwidth]{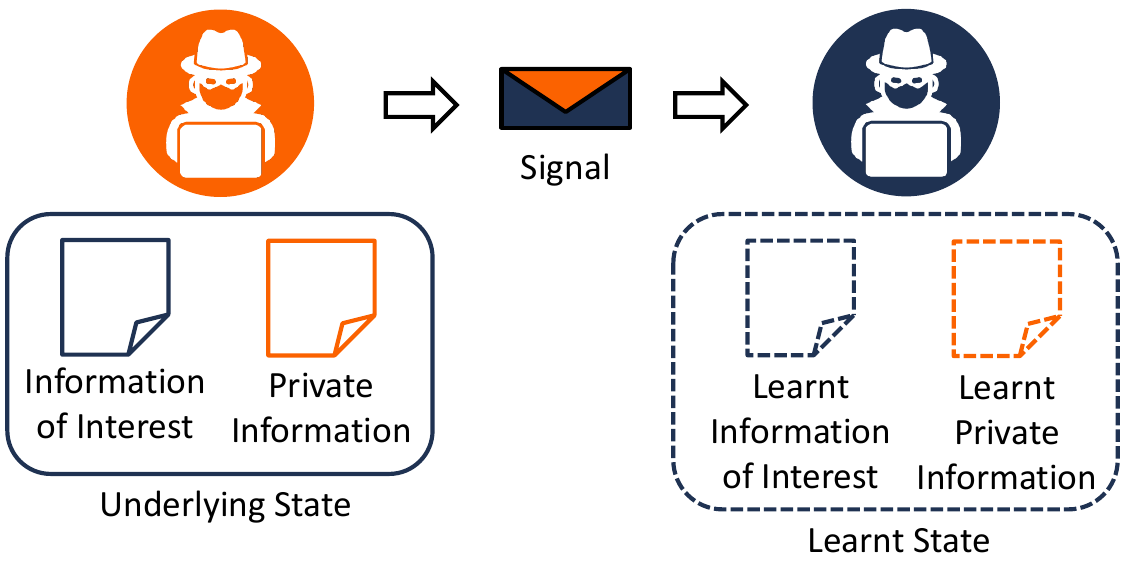}
\caption{An illustration of Bayesian persuasion with state-dependent quadratic cost measures when the state could consist of two types of information. Within the context of deception, the sender could seek to induce the receiver to perceive the information of interest as the private information. Within the context of privacy, the sender could seek to share the information of interest without leaking the private information since it might be inferred from the information of interest shared, due to the correlation between them.}\label{fig:model}
\end{figure}

There have also been computational approaches for the Bayesian persuasion problem, e.g., \cite{ref:Dughmi16,ref:Dughmi17}. Particularly, the non-strategic nature of the receiver makes it possible to formulate the problem as a single optimization problem faced by the sender. {\cb For scenarios where the receiver has a finite action space,} based on the revelation principle, the authors have formulated a linear program (LP)\footnote{This LP has a computational complexity polynomial with the size of the state space and the receiver's action space.} to compute the optimal signaling strategy, which, however, turns out to be impractical to solve numerically unless the finite state and action spaces are small \cite{ref:Dughmi16,ref:Dughmi17}. Therefore the authors consider the scenarios where the players' cost measures are independently (and identically) drawn from a known distribution for each state, and examine connections with auction theory. 

{\cb In this paper, our goal is to address Bayesian persuasion with state-dependent quadratic cost measures for general classes of distributions. More explicitly,} we consider the scenario where a sender has access to a realization (called state) of a random variable (with commonly known statistical profile) while a receiver seeks to compute Bayesian (possibly nonlinear) estimate of the state in the mean-square-error sense based on the signal he/she receives. However, the sender constructs the signal strategically in a stochastic way in order to manipulate the receiver's perception according to another state-dependent quadratic cost measure. This scheme, for example, could address optimal signaling in various scenarios within the context of deception and privacy, depending on how we configure the underlying state and state-dependent quadratic cost measures, as illustrated in Fig. \ref{fig:model}.

{\cb Since the receiver forms his/her perception according to the Bayesian estimate, the sender can compute his/her best signaling strategy by solving an optimization problem that incorporates the receiver's best response. However, quite contrary to the scenarios studied in \cite{ref:Dughmi16}, here the sender cannot exploit the revelation principle to turn the optimization problem he/she faces into a (finite) LP since the receiver's action set is the convex hull of the state space, i.e., it is a continuum even when the underlying state space is finite, except the trivial case when the state space is a singleton. Further, the optimization problem faced by the sender turns out to be non-convex and highly non-linear.}

One of our contributions in this paper is {\cb to exploit the quadratic structure of the cost measures in order} to transform the problem faced by the sender into a structured and exploitable form. To this end, for finite state spaces, we formulate an equivalent linear optimization problem over the convex cone of completely positive matrices\footnote{We say that a matrix $\Xi\in\universe{R}^{n\times n}$ is completely positive if there exists an (entry-wise) non-negative matrix $B\in\universe{R}^{n\times k}_+$ such that $\Xi = BB'$ \cite{ref:Berman03}.}. {\cb We further formulate a semi-definite relaxation of this equivalent problem, which can be solved effectively for relatively large state spaces. And the SDP-relaxation turns out to be tight if the size of the state space is less than or equal to $4$.} Furthermore we show that the strong dual of the equivalent problem is a linear optimization problem over the cone of copositive matrices\footnote{We say that a symmetric matrix $A\in\universe{S}^n$ is copositive provided that $\vec{b}' A \vec{b} \geq 0$ for all $\vec{b}\in\universe{R}_+^n$ \cite{ref:Berman03}.}. {\cb Therefore the sender could choose to solve either one while computing the optimal signaling strategy.} 

{\cb The equivalence relation established is of importance because} optimization over the cone of completely positive (or copositive) matrices is an active research area where the compact structure of the cones is approximated via polyhedral or semi-definite cones to any degree of accuracy at the expense of computational complexity, e.g., see \cite{ref:Parillo00, ref:deKlerk02, ref:Bundfuss09, ref:Yildirim12, ref:Yildirim17, ref:Bostanabad18}. {\cb To exemplify the effectiveness of the equivalence established, we examine the performance of sequential polyhedral approximations, provided by \cite{ref:Bundfuss09}, in the computation of optimal signaling. We further compare our solution concept with the LP formulation of Bayesian persuasion provided by \cite{ref:Dughmi16} (with a quantized version of the receiver's action space) and the SDP-relaxation formulated by \cite{ref:Tamura14} (which turns out to be tight for Gaussian information) in various scenarios numerically.}

The paper is organized as follows: In Section \ref{sec:problem}, we formulate the problem for the general class of distributions. {\cb In Section \ref{sec:main}, we provide the main equivalence result and discuss computational approaches. In Section \ref{sec:others}, we compare our approach to existing equivalent problem formulations.} In Section \ref{sec:examples}, we provide illustrative numerical examples. We conclude the paper in Section \ref{sec:conclusion}. An Appendix provides the proofs of the three technical results.

{\em Notation.} We denote random variables by bold lower case letters, e.g., $\rand{x}$. Sets are denoted by calligraphic letters, e.g., $\stateSet$. We denote the set of all probability distributions on a set $\stateSet$ by $\Delta(\stateSet)$. For a vector $x$ and a matrix $A$, $x'$ and $A'$ denote their transposes; further $\|x\|$ and $\|A\|_2$ denote the Euclidean ($L^2$) norms of the vector $x$ and the matrix $A$, respectively. For a matrix $A$, $\trace{A}$ denotes its trace. We denote the identity and zero matrices with the associated dimensions by $I$ and $O$, respectively. For positive semi-definite matrices $A$ and $B$, $A\succeq B$ means that $A-B$ is also a positive semi-definite matrix. $\universe{S}^m$ (or $\universe{S}^m_+$) denotes the set of symmetric (or positive semi-definite) matrices of dimensions $m$-by-$m$ while $\universe{R}^{n\times m}_+$ (or $\universe{R}^{n\times m}_{++}$) denotes the set of $n$-by-$m$ matrices with non-negative (or positive) entries. {\cb $\CP^n\subset \mathbb{R}^{n\times n}$ and $\COP^n\subset \mathbb{R}^{n\times n}$, respectively, denote the cone of completely positive and copositive matrices.}

\section{Problem Formulation} \label{sec:problem}

Consider two non-cooperating decision makers: a sender (\player{S}) and a receiver (\player{R}), as seen in Fig. \ref{fig:model}. {\cb \player{S} has access to an underlying state, which is realization of an $m$-dimensional random vector $\rand{\state}$ with (finite) support $\stateSet\subset\universe{R}^m$. Given the realization $z$, \player{S} sends a signal $\signal\in\signalsSet$ according to a stochastic kernel, denoted by $\Srule(\signal|z)$, i.e., $\Srule(\cdot|z)\in\Delta(\signalsSet)$ for each $z\in\stateSet$. Let us denote the set of all such signaling rules by $\Sspace$, i.e., $\Srule\in\Sspace$. On the other side, after a realization of the signal sent is received, e.g., $s\in\signalsSet$, \player{R} selects her\footnote{We use the pronouns ``he" and ``she" while referring to \player{S} and \player{R}, respectively, only for clear referral.} strategy $\Rrule(\cdot)$ that is a (Borel measurable) function from $\signalsSet$ to $\stateSet$ and her estimate of the underlying state is given by $\hat{\state} = \Rrule(\signal)$. The strategy space of \player{R} is denoted by $\Rspace$, which is the set of all (measurable) functions from $\signalsSet$ to $\stateSet$. We also define the random variable $\estimate{\state}:= \Rrule(\rand{\signal})$ almost everywhere over $\stateSet$.

The decision makers select their strategies according to distinct cost measures. \player{R} seeks for the mean-square-error estimate of the underlying state and has the cost measure\footnote{Note that the dependence of the estimate $\estimate{\state}$ on the signaling rule $\Srule\in\Sspace$ and the decision rule $\Rrule\in\Rspace$ is implicit for notational simplicity.}
\begin{equation}\label{eq:Rcost}
\cost{R}(\Srule,\Rrule) = \E{\|\rand{\state}-\estimate{\state}\|^2}
\end{equation}
to be minimized via $\Rrule\in\Rspace$. On the other side, \player{S} has the state-dependent quadratic cost measure
\begin{equation}\label{eq:Scost}
\cost{S}(\Srule,\Rrule) = \E{\begin{bmatrix} \rand{z} \\ \estimate{z}\end{bmatrix}'\begin{bmatrix}Q_{zz}&Q_{z\hat{z}}\\ Q_{\hat{z}z} & Q_{\hat{z}\hat{z}} \end{bmatrix}\begin{bmatrix} \rand{z} \\ \estimate{z} \end{bmatrix}},
\end{equation}
for some weight matrices $Q_{zz},Q_{z\hat{z}},Q_{\hat{z}z},Q_{\hat{z}\hat{z}}\in\universe{R}^{m\times m}$, to be minimized via $\Srule\in\Sspace$.

The flexibility on the weight matrices in \eqref{eq:Scost} enables us to model various problems within the same framework described above. For example, as illustrated in Fig. \ref{fig:model}, suppose that the underlying state consists of two types of information: information of interest $\info\in\infoSet$ and private information $\private\in\privateSet$, i.e., $\state =(\info,\private)$. Let us denote the estimates of these two types of information by $\hat{\info}$ and $\hat{\private}$, accordingly.

\begin{example}[Deception]
Consider the scenario where \player{S} seeks to induce \player{R} to perceive $\info$ as $\private$ while \player{R} seeks to estimate the true value of $\info$. Then, for this example, \player{S} could have the cost measure
\begin{equation}
\cost{S}(\Srule,\Rrule) = \E{\|\rand{\private}-\estimate{\info}\|^2},
\end{equation}
which could be obtained from \eqref{eq:Scost} provided that the weight matrices are set as
$Q_{zz} = \begin{psmallmatrix} O&O\\ O&I \end{psmallmatrix}$, $Q_{z\hat{z}}=Q_{\hat{z}z}' = \begin{psmallmatrix} O&O\\ -I&O \end{psmallmatrix}$, and  $Q_{\hat{z}\hat{z}} = \begin{psmallmatrix} I&O\\ O&O \end{psmallmatrix}$.
\end{example}

\begin{example}[Privacy]
Consider the scenarios where random variables $\rand{\info}$ and $\rand{\private}$ (that do not necessarily have the same dimensions) are not independent of each other, and \player{S} seeks to share $\info\in\infoSet$ while minimizing the informational leakage of $\private\in\privateSet$. Then for this example, \player{S} could have the cost measure
\begin{equation}
\cost{S}(\Srule,\Rrule) = \E{\|\rand{\info}-\estimate{\info}\|^2} - \E{\|\rand{\private}-\estimate{\private}\|^2},
\end{equation}
which could also be obtained from \eqref{eq:Scost} with the weight matrices set as
$Q_{zz} = \begin{psmallmatrix} I&O\\ O&-I \end{psmallmatrix}$, $Q_{z\hat{z}}=Q_{\hat{z}z}' = \begin{psmallmatrix} -I&O\\ O&I \end{psmallmatrix}$, and $Q_{\hat{z}\hat{z}} = \begin{psmallmatrix} I&O\\ O&-I \end{psmallmatrix}$.
\end{example}

Under the solution concept of Stackelberg equilibrium \cite{ref:Basar99}, policy commitment of \player{S} implies that \player{S} is the leader while \player{R} is the follower. Let $B(\Srule) \subset \Rspace$ be the optimum reaction set of the follower \player{R} corresponding to a given strategy $\Srule\in\Sspace$ of \player{S}. Note that $B(\Srule)$ is actually a singleton for any given signaling strategy $\Srule\in\Sspace$ since the best reaction of \player{R} is unique and given by $\hat{\state} = \E{\rand{\state}|\rand{\signal}=\signal}$ (and correspondingly $\estimate{z}=\E{\rand{\state}|\rand{\signal}}$ almost everywhere over $\stateSet$). Then, the pair of the strategy and the optimum reaction, $(\Srule^*,B(\Srule^*))$, attains the Stackelberg equilibrium provided that
\begin{subequations}\label{eq:SE}
\begin{align}
&\Srule^* \in\argmin_{\Srule\in\Sspace} \cost{S}(\Srule,B(\Srule^*))\\
&B(\Srule) = \argmin_{\Rrule\in\Rspace} \cost{R}(\Srule,\Rrule).
\end{align}
\end{subequations}}

\section{Main Result} \label{sec:main}

{\cb In this section, we provide a solution concept for the computation of optimal signaling strategies that attain the Stackelberg equilibrium, as described in \eqref{eq:SE}.} Before delving into the technical details, let us first provide an overview of the main result of this paper: We start by identifying the optimization problem faced by \player{S} given the non-strategic machinery of Bayesian estimation followed by \player{R}. {\cb Then we show that the optimization problem turns out to be highly nonlinear and non-convex, which poses a challenge for solving it globally. To mitigate this challenge,} we formulate an equivalent linear optimization problem over the cone of completely positive matrices {\cb and show that its (strong) dual is a linear optimization problem over the cone of copositive matrices}. 

{\cb From a computational perspective, we formulate an SDP relaxation of the equivalent problem and analyze its computational complexity. We further provide one computational approach adopted from the literature on copositive programming to solve the optimal signaling problem effectively. Finally we quantify the approximation level of adopting our solution concept for a quantized version of a continuum state space.}

In the following, we provide the technical details.

\subsection{Optimization Problem Faced by \player{S}}
{\cb Since the best reaction of \player{R} is uniquely given by $\estimate{\state} = \E{\rand{\state}|\rand{\signal}}$ almost everywhere over $\stateSet$, the optimization problem faced by \player{S}, as defined in \eqref{eq:Scost}, can be written as
\begin{align}
\min_{\Srule\in\Sspace}\cost{S}(\Srule,\Rrule^*) &= \min_{\Srule\in\Sspace}\left[\E{\rand{z}'Q_{zz}\rand{z}} + \E{\estimate{z}'V\estimate{z}}\right]\label{eq:ScostIncorp}\\
&=\E{\rand{z}'Q_{zz}\rand{z}} + \min_{\Srule\in\Sspace}\E{\estimate{z}'V\estimate{z}}\label{eq:Sfaces}
\end{align}
where 
\begin{equation}\label{eq:V}
V := Q_{z\hat{z}}+Q_{\hat{z}z} + Q_{\hat{z}\hat{z}}
\end{equation}
since we have $\E{\rand{z}\estimate{z}'} = \E{\E{\rand{z}\estimate{z}'|\rand{s}}}= \E{\estimate{z}\estimate{z}'}$. And \eqref{eq:Sfaces} follows since the first term on the right-hand-side of \eqref{eq:ScostIncorp} does not depend on the optimization argument $\Srule\in\Sspace$.

\begin{example}[Deception - Cont'd]
Within the context of deception, the weight matrix in \eqref{eq:ScostIncorp} is given by $V=\begin{psmallmatrix} I&-I\\-I&O\end{psmallmatrix}$.
\end{example}

\begin{example}[Privacy - Cont'd]
Within the context of privacy, the weight matrix in \eqref{eq:ScostIncorp} is given by $V=\begin{psmallmatrix} -I&O\\O&I\end{psmallmatrix}$.
\end{example}}

Suppose that the prior distribution has complete support on $\stateSet$, $n:=|\stateSet|$ and let $\prior(z)\in(0,1]$ denote the probability that the state $z\in\stateSet$ is realized. Then Bayes rule yields that the probability of state $z\in\stateSet$ being realized given that signal $\signal\in\signalsSet$ is received is given by\footnote{A signal $\signal\in\signalsSet$ would be received if the associated probability is positive, i.e., $p(\signal)>0$.}
\begin{equation}\label{eq:posterior}
\posterior(z) := \frac{\Srule(\signal|z)\prior(z)}{p(\signal)},
\end{equation}
where $p(s)$, denoting the probability that the signal $\signal\in\signalsSet$ is sent, is given by
\begin{equation}\label{eq:signalProb}
p(s) = \sum_{z\in\stateSet} \Srule(\signal|z)\prior(z).
\end{equation}
Therefore, for given $\signal\in\signalsSet$, we have\footnote{\cb This shows that \player{R}'s action set, i.e., the set of all possible posterior estimates of the state, is the convex hull of $\stateSet$, i.e., a continuum unless $\stateSet$ is a singleton.}
\begin{align}
\hat{z}:=\E{\rand{z}|\rand{\signal}=\signal}= \sum_{z\in\stateSet} \posterior(z) z.
\label{eq:meanPosterior}
\end{align}
{\cb Correspondingly, the optimization objective of \player{S}, as described in \eqref{eq:Sfaces}, could be written as
\begin{align}
&\E{\estimate{\state}'V\estimate{\state}} = \sum_{\signal\in\signalsSet} p(\signal) \left(\sum_{\state\in\stateSet}\posterior(\state)\state\right)'V\left(\sum_{\state\in\stateSet}\posterior(\state)\state\right),\\
&=\sum_{\signal\in\signalsSet} \frac{\left(\sum_{\state\in\stateSet}\Srule(\signal|\state)\prior(\state)\state\right)'V\left(\sum_{\state\in\stateSet}\Srule(\signal|\state)\prior(\state)\state\right)}{\sum_{\state\in\stateSet}\Srule(\signal|\state)\prior(\state)},\label{eq:covariance}
\end{align}
which is, in general, a highly nonlinear and nonconvex function of the optimization argument $\Srule\in\Sspace$ (due to the denominator term in \eqref{eq:covariance}). Since a local solution can be arbitrarily bad in a non-convex optimization problem, next we formulate an equivalent problem for which a {\em global} optimum could be computed via existing solvers effectively.}

\subsection{\cb An Equivalent Linear Optimization Problem}

{\cb Note that $\E{\estimate{\state}'V\estimate{\state}} = \trace{\E{\estimate{z}\estimate{z}'}V}$ and} we can write the correlation matrix of the posterior state $\E{\estimate{z}\estimate{z}'}$ in a compact form as
\begin{equation}
\E{\estimate{z}\estimate{z}'} = Z\myXi Z',
\end{equation}
where $Z := \begin{bmatrix} z_1 \ldots z_n\end{bmatrix}\in\universe{R}^{m\times n}$ and we introduce $\myXi \in\universe{S}^n$ whose $i$th row and $j$th column entry is given by
\begin{equation}\label{eq:entry}
\myXi[i,j] = \sum_{\signal\in\signalsSet} p(\signal) \posterior(z_i) \posterior(z_j).
\end{equation}
Then, the problem faced by \player{S} can be written as
\begin{align}\label{eq:XiEquivalent}
\min_{\Srule\in\Sspace} \,\trace{\E{\estimate{z}\estimate{z}'}V} = \min_{\Srule\in\Sspace} \,\trace{\myXi\bar{V}},
\end{align}
where $\bar{V} := Z'VZ$. 

The following proposition provides a {\cb concrete} characterization of a necessary and sufficient condition on $\myXi$.

\begin{proposition}\label{prop:main}
For any signaling rule $\Srule\in\Sspace$, the symmetric matrix $\myXi$ satisfies $\myXi\in\CP^n$ and $\myXi\vec{1} = \vec{p}_o$.\footnote{We define $\vec{p}_o := \begin{bmatrix} \prior(z_1) &\ldots&\prior(z_n)\end{bmatrix}'$.} Furthermore for any completely positive matrix $\Xi$ that satisfies $\Xi\vec{1} = \vec{p}_o$, there exists a signaling rule $\Srule\in\Sspace$ such that $\myXi = \Xi$. 

Consider a factorization\footnote{A completely positive matrix could have multiple different factorizations even for the same $k\in\universe{N}$ \cite{ref:Dickinson12}. This also yields that the corresponding signaling strategy is not unique in general.} of the completely positive matrix $\Xi$, as $\Xi^* = \sum_{i=1}^k \vec{b}_i\vec{b}_i'$, where $\vec{b}_i\in\universe{R}_+^{n}$. Then, the corresponding signaling strategy is given by\footnote{Note that the prior distribution has complete support over $\stateSet$ by definition.}
\begin{equation}\label{eq:optimalSignal}
\Srule^*(\signal_i|z_j) = \vec{b}_i' \vec{1} \frac{b_{i,j}}{\prior(z_j)},
\end{equation}
for all $i=1,\ldots,k$ and $j=1,\ldots,n$, where $b_{i,j}\geq 0$ denotes the $j$th entry of the vector $\vec{b}_i$.
\end{proposition}

\begin{myproof}
The proof is provided in Appendix \ref{app:main}.
\end{myproof}

Based on Proposition \ref{prop:main}, the following corollary shows that the problem faced by \player{S} could be written, ``equivalently", as a linear optimization problem over the convex cone of completely positive matrices. 

\begin{corollary}\label{corollary:main}
The problem faced by \player{S} can be written in an equivalent form as
\begin{align}\label{eq:CPequivalent}
\min_{\Srule\in\Sspace} \,\trace{\E{\estimate{z}\estimate{z}'}V} = \min_{\Xi\in\CP^n}& \;\trace{\Xi\bar{V}},\\
\subjectTo& \;\Xi\vec{1} = \vec{p}_o\nn
\end{align}
where $\bar{V} = Z'VZ$.
\end{corollary}

{\cb The following proposition shows that} given the primal problem:
\begin{align}\label{eq:primal}\tag{P}
\min_{\Xi\in\CP^n} \;\trace{\Xi\bar{V}}, \subjectTo& \;\Xi\vec{1} = \vec{p}_o,
\end{align}
{\cb its {\em strong} dual (with respect to trace inner product)} is given by
\begin{align}\label{eq:dual}\tag{D}
\max_{\vec{y}\in\universe{R}^n,S\in\COP^n} \;\vec{p}_o' \vec{y}, \subjectTo\; \vec{1}\vec{y}' + S = \bar{V}.
\end{align}
{\cb This enables us to solve either of them while obtaining the value of both.} 

\begin{proposition}\label{prop:duality}
{\cb The dual of \eqref{eq:primal} with respect to trace inner product is given by \eqref{eq:dual}. Furthermore,} \eqref{eq:primal} is feasible, has finite value, and has an interior point. Therefore, there exists a strong duality between \eqref{eq:primal} and \eqref{eq:dual}, i.e., {\cb we always have $\eqref{eq:primal} = \eqref{eq:dual}$}.
\end{proposition}

\begin{myproof}
The proof is provided in Appendix \ref{app:duality}.
\end{myproof}

In the following subsections, we examine the equivalent problem formulation from a computational perspective.

\subsection{Semi-definite Program Relaxation}\label{sec:SDP}

Note that $\CP^n \subseteq \universe{S}_+^n \cap \universe{R}_+^{n\times n}$ for all $n$, while $\CP^n = \universe{S}_+^n \cap \universe{R}_+^{n\times n}$ if, and only if, $n\leq 4$ \cite{ref:Maxfield62,ref:Berman03}. {\cb Based on this fact, the following proposition provides an SDP relaxation of \eqref{eq:primal}.

\begin{proposition}\label{prop:interesting}
An SDP relaxation of the primal problem \eqref{eq:primal} is given by
\begin{align}\label{eq:SDP}\tag{SDP}
\min_{\Xi\in\universe{S}_+^n} \;\trace{\Xi\bar{V}},\subjectTo \;\Xi\vec{1} = \vec{p}_o, \Xi \in\universe{R}_+^{n\times n},
\end{align}
and we always have $\eqref{eq:primal}\geq\eqref{eq:SDP}$, with equality if $n\leq 4$.
\end{proposition}

Due to the constraints $\Xi\vec{1} = \vec{p}_o$ and $\Xi \in\universe{R}_+^{n\times n}$, it can be shown that the maximum Frobenius norm of any feasible solution of the SDP relaxation is bounded from above by $1$. Based on this observation and \cite[Theorem 2.6.1]{ref:Gartner12}, the following corollary to Proposition \ref{prop:interesting} characterizes the computational complexity of solving the SDP relaxation.

\begin{corollary}
There exists an algorithm that can solve \eqref{eq:SDP} up to any desired level of accuracy with run-time complexity polynomial in the size of the state space.
\end{corollary}}

\subsection{\cb Adopting Copositive Programming Methods}\label{sec:computation}

{\cb Based on the equivalence results provided above, we can adopt existing computational tools that have been developed for copositive programming in the optimization community, e.g., \cite{ref:Parillo00, ref:deKlerk02, ref:Bundfuss09, ref:Yildirim12, ref:Yildirim17, ref:Bostanabad18}.} Furthermore, each new development in this active research area (due to its broad applications) will bring in new computational tools and insights to address the optimal signaling problem. 

For example, we can approximate $\CP^n$ through polyhedral cones \cite{ref:Bundfuss09, ref:Yildirim12, ref:Yildirim17} from inside and outside of $\CP^n$ such that their (nested) sequence converges to $\CP^n$ asymptotically. For each polyhedral approximation, the corresponding optimization problem becomes an LP, which can be solved in polynomial time. A comparison between inner and outer approximations quantifies the accuracy of the approximation.
{\cb Furthermore, in \cite{ref:Bundfuss09}, the authors have proposed an algorithm that attains fine approximation only around some neighborhood of the solution of the optimization problem while letting the approximation be coarse anywhere else in order to speed it up. In the following we briefly describe the algorithm since we will adopt it to solve \eqref{eq:primal} and examine its performance numerically in various scenarios in Section \ref{sec:examples}.}

Note that the extreme rays of $\CP^n$ have rank $1$, i.e., they can be written as $\vec{b}\vec{b}'$, where $\vec{b}\in\universe{R}_+^n$ \cite{ref:Berman03}. Correspondingly, $\CP^n$ can be viewed as the conic hull of vectors from the unit simplex $\Delta_{n-1}$ in $\universe{R}^n$ \cite{ref:Yildirim17}. Consider a family of simplices $\set{P} = \{\Delta^1,\ldots,\Delta^t\}$ satisfying
\begin{equation*}
\Delta_{n-1} = \bigcup_{i=1}^t \Delta^i \mbox{ and } \interior{\Delta^i}\,\cap\, \interior{\Delta^j} = \varnothing \mbox{ if } i\neq j.
\end{equation*}
Given the family of simplices $\set{P}$, we define the polyhedral cones:\footnote{The subscripts $b$ and $c$ denote the index of the associated vectors.}
\begin{align}
\set{I}_{\set{P}} &:= \left\{\sum_{\vec{b}\in V_{\set{P}}} \lambda_{b}\vec{b}\vec{b}' : \lambda_b \geq 0\right\},\label{eq:inner}\\
\set{O}_{\set{P}} &:= \left\{\sum_{\vec{b},\vec{c}\in V_{\set{P}}} \lambda_{b,c}(\vec{b}\vec{c}'+\vec{c}\vec{b}') : \lambda_{b,c} \geq 0\right\},\label{eq:outer} 
\end{align}
where $V_{\set{P}}$ denotes the set of vertices in $\set{P}$. In \cite{ref:Bundfuss09}, the authors have shown that $\set{I}_{\set{P}}\subseteq \CP^n \subseteq \set{O}_{\set{P}}$ for any $\set{P}$. Let $\mathrm{CP}(\set{K})$ denote the solution of the following $\set{K}$-cone program:
\begin{align}\label{eq:CPK}
\mathrm{CP}(\set{K}) = \min&\;\trace{\Xi\bar{V}}\\
\subjectTo &\;\Xi\vec{1} = \vec{p}_o\nn\\
&\;\Xi\in\set{K}\nn
\end{align}
Since $\set{I}_{\set{P}}\subseteq \CP^n \subseteq \set{O}_{\set{P}}$, we have
\begin{equation}\label{eq:UpperBelow}
\mathrm{CP}(\set{I}_{\set{P}}) \geq \mathrm{CP}(\CP^n) \geq \mathrm{CP}(\set{O}_{\set{P}}).  
\end{equation}
Correspondingly, through a sequence of simplical partitions, we can construct a sequence of nested polyhedral cones $\set{I}_1\subseteq \set{I}_2\subseteq \ldots$ and $\set{O}_1\supseteq \set{O}_2 \supseteq \ldots $ that converge to $\CP^n$, i.e.,
\begin{equation}
\CP^n = \overline{\bigcup_{i\in\mathbb{N}} \set{I}_i} \mbox{ and }\CP^n = \bigcap_{i\in\mathbb{N}} \set{O}_i,
\end{equation}
from below and above, respectively \cite{ref:Bundfuss09}.

\subsection{Approximation Guarantees for Continuum State Spaces}\label{sec:quanta}

{\cb Up to now we have focused on finite state spaces. For continuum state spaces, we can still adopt the solution concept accompanied with a quantization scheme. The following corollary to Proposition \ref{prop:main} quantifies the associated approximation error in terms of the quantization error. Note that a smaller quantization error can necessitate large number of quantization bins, and therefore the associated problem might have a large state space. In such cases, the error quantified can also be incorporated into the SDP-relaxation since it can be solved via existing SDP solvers effectively for relatively large state spaces.}

\begin{corollary}\label{corollary:quantization}
Consider a quantization of the continuous random variable $\rand{z}\in\stateSet$, denoted by $\rand{z}_q\in\stateSet$, i.e., $\rand{z}_q$ attains the same value within any bin of the quantization. Let $\rand{e}=\rand{z}-\rand{z}_q$, almost everywhere over $\stateSet$, denote the quantization error. Then, we have\footnote{By turning the problem around, we can view it as \player{S} selecting a random vector within the general class of square integrable distributions and sending a realization of that signal rather than selecting a signaling strategy. Note that \player{S} should take into account the joint distribution of the underlying distribution and the signal sent, which would normally have been determined by the signaling strategy.}
\begin{align}
\left|\min_{\rand{\signal}} \;\trace{\E{\estimate{z}\estimate{z}'}V} - \min_{\rand{\signal}} \trace{\E{\estimate{z}_q\estimate{z}_q'}V}\right| \leq \epsilon,
\end{align}
where $\estimate{z} = \E{\rand{z}|\rand{\signal}}$, $\estimate{z}_q = \E{\rand{z}_q|\rand{\signal}}$, and
\begin{equation}
\epsilon = (2\|\rand{z}_q\| + \|\rand{e}\|)\|V\|_2\|\rand{e}\|,\label{eq:epsilon}
\end{equation}
which yields that $\epsilon\rightarrow 0$ when $\|\rand{e}\|\rightarrow 0$. {\cb Furthermore, let $n_q$ denote the number of bins in the quantization scheme. Then, we also have
\begin{equation*}
\min_{\Srule\in\Sspace} \trace{\E{\estimate{z}\estimate{z}'}V} + \epsilon \geq \begin{array}{rl}\min_{\Xi\in\universe{S}_+^{n_q}} & \trace{\Xi\bar{V}}\\
\subjectTo &\Xi\vec{1} = \vec{p}_o, \Xi \in\universe{R}_+^{n_q\times n_q}.\end{array}
\end{equation*}}
\end{corollary}

\begin{myproof}
The proof is provided in Appendix \ref{app:quantization}.
\end{myproof}

\section{\cb Other Equivalent Problem Formulations}\label{sec:others}

{\cb In this section, we compare our equivalent problem formulation with other equivalent problems provided by \cite{ref:Kamenica11}, \cite{ref:Dughmi16}, and \cite{ref:Tamura14} (in this order). Note that \cite{ref:Kamenica11} and \cite{ref:Dughmi16} are for general classes of cost measures, whereas ours is {\em exclusive} to state-dependent quadratic cost measures. Although  \cite{ref:Tamura14} is also for state-dependent quadratic cost measures, it is specific to Gaussian distributions whereas ours is for any distribution (with finite support). 

\subsection{A Continuous LP Based on Bayes Plausibility}

In \cite{ref:Kamenica11}, the authors have shown that the sender could induce a posterior belief $p$ on the underlying state (or information) via a signaling strategy if, and only if, the posterior belief is {\em Bayes plausible}, i.e., $\int_{\Delta(\stateSet)} p \tau(dp) = p_o$, where $\tau$ denotes the probability measure over the posterior beliefs induced by the signaling strategy. Correspondingly if we let $\hat{c}_S(p)$ correspond to the cost of \player{S} for the posterior belief $p$, then the optimization problem faced by \player{S} is equivalent to 
\begin{equation}\label{eq:BP}
\min_{\tau\in\Delta(\Delta(\stateSet))} \int_{\Delta(\stateSet)} \hat{c}_S(p)\tau(dp)\;\subjectTo  \int_{\Delta(\stateSet)} p\tau(dp) = \prior.
\end{equation}
In other words, \player{S} can look for $\tau$, a distribution over posterior beliefs, in order to compute the best signaling strategy. However $\tau$ is defined over $\Delta(\Delta(\stateSet))$, which is an infinite dimensional space even for finite state spaces, where $|\stateSet|<\infty$.

Alternatively, in order to solve the equivalent problem \eqref{eq:BP}, \cite{ref:Kamenica11} proposes a solution concept that uses convexification. In particular, let $C(p)$ be the convex envelope of $\hat{c}_S(p)$, defined by
\begin{equation}
C(p) \equiv \inf\{c | (p,c)\in \co{\hat{c}_S}\},
\end{equation} 
where $\co{\hat{c}_S}\subset \Delta(\stateSet)\times \mathbb{R}$ denotes the convex hull of the graph of $\hat{c}_S$. Then its convex nature yields that the minimum cost for \player{S} is simply equal to $C(\prior)$. However, this necessitates the computation of a convex envelope in $|\stateSet|$-dimensional space, which is challenging for $|\stateSet|>2$, except the trivial cases where $\hat{c}_S$ turns out to be a convex or a concave function. 

In our problem setting, the cost for \player{S} as a function of the posterior belief $p$ turns out to be {\em neither convex nor concave} in general. Particularly, it is given by
\begin{equation}
\hat{c}_S(\vec{p}) = \trace{Z'VZ \vec{p}\vec{p}'},
\end{equation}  
and Sylvester's law of inertia yields that $Z'VZ$ is not necessarily a definite matrix since $V$ is not necessarily a definite matrix, as exemplified in deception and privacy scenarios. In other words, if \player{S} adopted the solution concept of \cite{ref:Kamenica11} to compute his signaling strategy, then in general he would have faced the non-trivial cases where he needs to compute the convex envelope.

The following remark establishes a connection between \eqref{eq:BP} and the polyhedral approximations we adopted in Subsection \ref{sec:computation}.}

\begin{remark}
{\cb Based on the definition of inner polyhedral approximation $\set{I}_{\set{P}}$, \eqref{eq:inner}, and the description of $\mathcal{K}$-cone program, \eqref{eq:CPK}; $\mathrm{CP}(\set{I}_{\set{P}})$ is given by}
\begin{align}
\min_{\vec{\lambda}\in\universe{R}_+^{|V_{\set{P}}|}} \; \trace{\sum_{\vec{b}\in V_{\set{P}}}\lambda_b \vec{b}\vec{b}' \bar{V}} \;\subjectTo \; \sum_{\vec{b}\in V_{\set{P}}}\lambda_b \vec{b}(\vec{b}' \vec{1}) = \vec{p}_o,
\end{align}
which is equivalent to
\begin{align}\label{eq:BPlike}
\min_{\lambda\in\universe{R}_+^{|V_{\set{P}}|}} \sum_{\vec{b}\in V_{\set{P}}} c_b\lambda_b \;\subjectTo \sum_{\vec{b}\in V_{\set{P}}} \vec{b} \lambda_b= \vec{p}_o,
\end{align}
where $c_b := \trace{\vec{b}\vec{b}'\bar{V}}$.
The resemblance between \eqref{eq:BP} and \eqref{eq:BPlike} is notable. Particularly, a discretization of the simplex $\Delta(\stateSet)$ in \eqref{eq:BP} would have led to \eqref{eq:BPlike}. {\cb This yields that if we just seek to solve \eqref{eq:BP} directly via an LP solver after discretizing the simplex $\Delta(\stateSet)$, then it could approximate the optimal signaling cost only from above and it would not be possible to quantify the approximation error and adapt the quality of polyhedral approximations without the outer polyhedral approximation described in Subsection \ref{sec:computation},  which plays an important role in speeding up the algorithm.}
\end{remark}

\subsection{\cb A (Continuous) LP Based on Incentive-Compatibility}\label{sec:LP}

{\cb Suppose that \player{R} has a finite action set (denoted by $\mathcal{U}$, i.e., $|\mathcal{U}|<\infty$), even though it is not the case in our problem setting. For such cases, in \cite{ref:Dughmi16}, the authors have made the observation that without loss of generality \player{S} could adopt a direct recommendation scheme in which he recommends only {\em incentive-compatible} actions $u\in\mathcal{U}$ to \player{R} such that \player{R}'s best response is to always follow the recommendation. 

Let $c_S(u,z)$ and $c_R(u,z)$ denote the costs for \player{S} and \player{R}, respectively, when the underlying state is $z$ and \player{R} takes the action $u$. Correspondingly, a signal $u$ is incentive-compatible provided that
\begin{equation}\label{eq:IC}
\sum_{z\in\stateSet} \prior(z) \pi(u|z) c_R(u,z) \leq \sum_{z\in\stateSet} \prior(z) \pi(u|z) c_R(\tilde{u},z) \tag{IC}
\end{equation}
for all $\tilde{u}\in\mathcal{U}$. Then the optimization problem faced by \player{S} could be written as
\begin{align}\label{eq:LLP}
\begin{array}{ll}
\min_{\pi\in\Pi} & \sum_{z\in\stateSet} \sum_{u\in\mathcal{U}} \prior(z)\pi(u|z)c_S(u,z)\\
\subjectTo & \sum_{u\in\mathcal{U}} \pi(u|z)ds = 1,\; \forall\; z\in\stateSet\\
& \pi(u|z)\geq 0,\; \forall\; z\in\stateSet\mbox{ and } u\in\mathcal{U}\\
& \eqref{eq:IC} \mbox{ holds for all }u,\tilde{u}\in\mathcal{U}.
\end{array}
\end{align}

Note that \eqref{eq:LLP} is an LP whose numbers of optimization arguments and constraints are polynomial in the size of the state space and \player{R}'s action space, i.e., $|\stateSet|$ and $|\mathcal{U}|$, respectively. This yields that there exists an algorithm that can solve it with run-time complexity polynomial in the size of the state space and \player{R}'s action space. Since \player{R}'s action space is a continuum in our problem setting (particularly it is the convex hull of the state space as can be seen in \eqref{eq:signalProb}), here it is not possible to adopt \eqref{eq:LLP} directly to compute optimal signaling strategies even for small-size finite state spaces. 

For example, if we extend \eqref{eq:LLP} to scenarios where \player{R} has a continuum action set (e.g., by defining a proper measurable signaling strategy and incentive-compatibility constraint, which necessitates more involved analysis), then \eqref{eq:LLP} turns into a continuous LP. Alternatively, let us also consider the scenarios where boundedly rational \player{S} quantizes the continuum action set of \player{R}, e.g., uniformly, to a finite number of quanta to adopt \eqref{eq:LLP} while computing his signaling strategy. However, smaller approximation error necessitates excess number of quanta, which corresponds to excessive computational load to solve the associated LP even when the state space is finite and small. Furthermore, the associated LP does not bound the original optimization problem from below since the relaxation is not with respect to the constraints on the optimization arguments quite contrary to the cases in the SDP-relaxation and the quantization of the continuum state space examined in Subsections \ref{sec:SDP} and \ref{sec:quanta}, respectively.

\subsection{An SDP for Gaussian Distributions}\label{sec:Gauss}

{\cb Let us take another look at the optimization problem faced by \player{S} when players have state-dependent quadratic cost measures, as described in \eqref{eq:Sfaces}, which shows that the signaling strategy has an impact on the optimization objective only through the correlation matrix of the posterior estimate, $\E{\estimate{z}\estimate{z}'}$. In \cite{ref:Tamura14}, the author has identified the necessary conditions on $\E{\estimate{z}\estimate{z}'}$ as
\begin{align}
\E{\rand{z}\rand{z}'} \succeq \E{\estimate{z}\estimate{z}'} \succeq  \E{\rand{z}}\E{\rand{z}}',\label{eq:necesTamura}
\end{align}
where the upper (or lower) bound corresponds to the scenario where \player{S} sends full (or null) signal. Note that \eqref{eq:necesTamura} is a linear matrix inequality while the optimization objective in \eqref{eq:Sfaces} is a linear function of $\E{\estimate{z}\estimate{z}'}$. Correspondingly, \cite{ref:Tamura14} has presented an SDP relaxation of \eqref{eq:Sfaces} given by
\begin{align}\label{eq:cc}
\begin{array}{l}\min_{S\in\universe{S}_+ ^n} \trace{SV}\\
\subjectTo\; \E{\rand{z}\rand{z}'} \succeq S \succeq  \E{\rand{z}}\E{\rand{z}}'.\end{array}
\end{align}
This relaxation \eqref{eq:cc} turns out to be tight if $\rand{z}$ is a (multivariate) Gaussian random variable, as shown in \cite{ref:Tamura14}, whereas it is not tight in general for other distributions, as shown in \cite{ref:Sayin19c}. Furthermore, \eqref{eq:cc} depends only on the second-order statistics of the underlying distribution whereas our SDP-relaxation depends on the probability distribution and is tight for state spaces of size $4$ or less, independent of the underlying distribution.}

In the following section, we compare the performance of these solution concepts with ours over various numerical examples.}

\begin{figure*}[t!]
     \centering
     \begin{subfigure}[b]{0.32\textwidth}
         \centering
         \includegraphics[width=\textwidth]{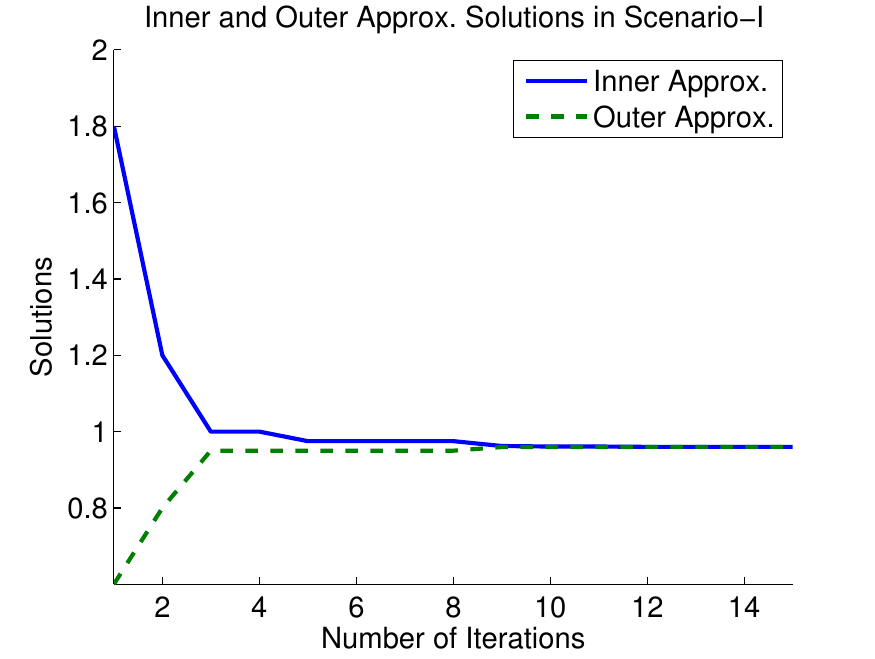}
         \caption{Scenario I}
         \label{fig:I}
     \end{subfigure}
     \hfill
     \begin{subfigure}[b]{0.32\textwidth}
         \centering
         \includegraphics[width=\textwidth]{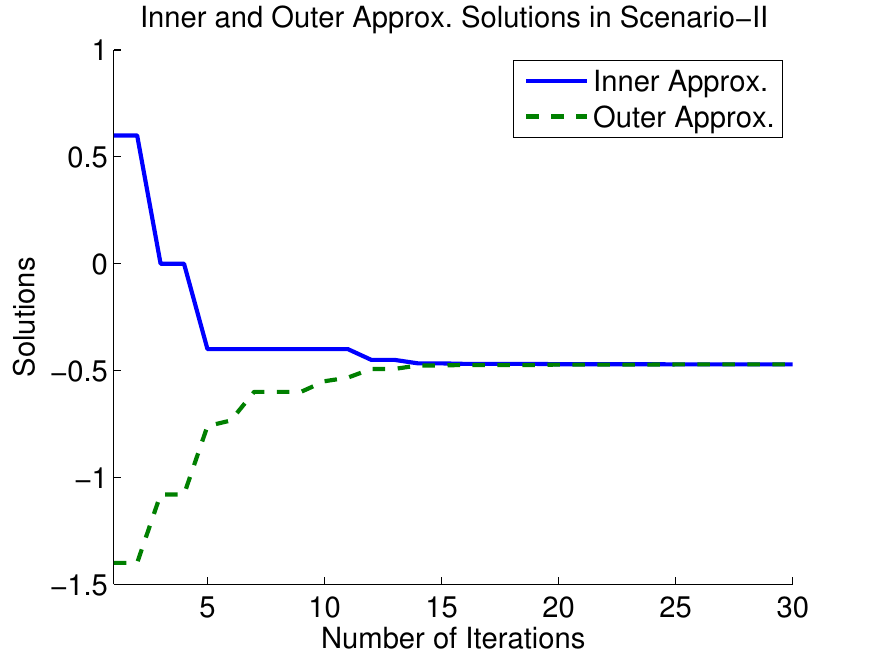}
         \caption{Scenario II}
         \label{fig:II}
     \end{subfigure}
     \hfill
     \begin{subfigure}[b]{0.32\textwidth}
         \centering
         \includegraphics[width=\textwidth]{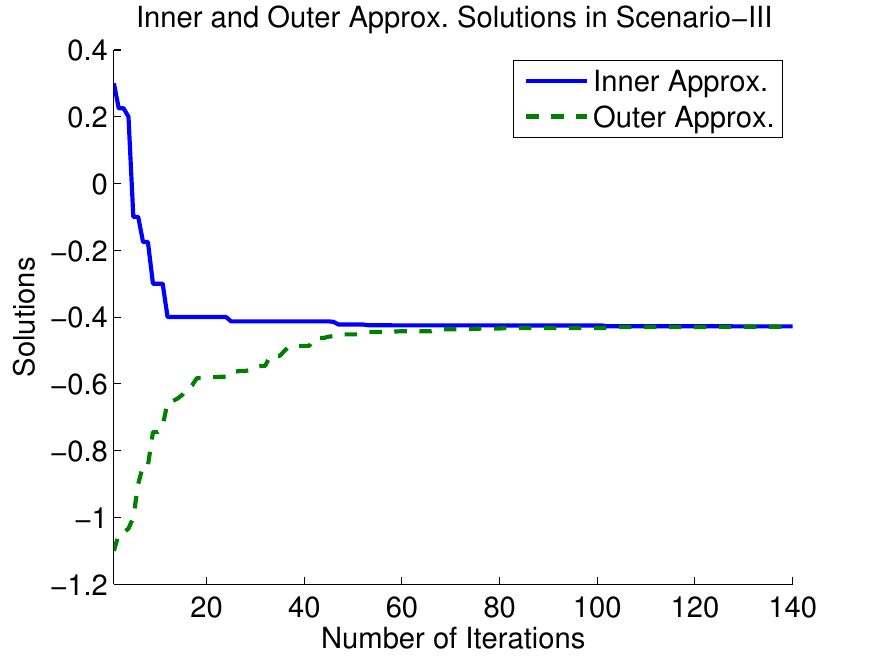}
         \caption{Scenario III}
         \label{fig:III}
     \end{subfigure}
        \caption{Computation of the minimum cost for \player{S} via sequential inner and outer polyhedral approximations within the context of deception. {\cb Similar convergence behavior is observed within the context of privacy.}}
        \label{fig:deception}
\end{figure*}

\section{Illustrative Examples}\label{sec:examples}

Similar to the example introduced in \cite{ref:Kamenica11}, let us consider the interaction between a prosecutor (\player{S}) and a judge (\player{R}) during a trial of a defendant. Particularly, now the prosecutor has access to 
\begin{itemize}
\item the information of interest $\info\in\infoSet$ corresponding to the status of a defendant based on some evidence
\item the private information $\private\in\privateSet$ corresponding to the prosecutor's intuition about the status of the defendant
\end{itemize}
Irrespective of the evidence, our self-confident and righteous prosecutor seeks to induce the judge to perceive the status of the defendant in line with his intuition. On the other side, the judge is only interested in what the evidence says about the status of the defendant. We first consider three scenarios where the underlying joint distributions over the prosecutor's intuition and what the evidence suggests about the defendant's status are {\cb chosen arbitrarily} as tabulated in Tables \ref{tab:joint1}, \ref{tab:joint2}, and \ref{tab:joint3}, where (I), (NG), and (G) correspond to the status of `innocent', `not guilty', and `guilty', respectively. {\cb We will examine our solution concept in a relatively larger state space in Scenario IV later in this section.} 

\begin{table}[t!]
\renewcommand{\arraystretch}{1.5}
\caption{Joint distribution in Scenario I.}\label{tab:joint1}
\begin{center}
\begin{tabular}{ m{4em}c|c|c} 
\multicolumn{2}{r}{\multirow{2}{*}{Scenario I. }} &\multicolumn{2}{c}{Evidence Suggests} \\
\multicolumn{2}{c|}{} & (G) $\rand{x} = -1$ & (I) $\rand{x} = 1$ \\ 
\cline{2-4}
\shortstack{Prosecutor\\Thinks} & (G) $\rand{y}=-1$ & 0.3 & 0.7 
\end{tabular}
\end{center}
\end{table}

\begin{table}[t!]
\renewcommand{\arraystretch}{1.5}
\caption{Joint distribution in Scenario II.}\label{tab:joint2}
\begin{center}
\begin{tabular}{ m{4em}c|c|c} 
\multicolumn{2}{r}{\multirow{2}{*}{Scenario II. }} &\multicolumn{2}{c}{Evidence Suggests} \\
\multicolumn{2}{c|}{} & (G) $\rand{x} = -1$ & (I) $\rand{x} = 1$ \\ 
\cline{2-4}
\multirow{2}{*}{\shortstack{Prosecutor\\Thinks}} & (I) $\rand{y}=1$ & 0.1 & 0.4 \\ 
\cline{2-4}
& (G) $\rand{y}=-1$ & 0.2 & 0.3
\end{tabular}
\end{center}
\end{table}

\begin{table}[t!]
\renewcommand{\arraystretch}{1.5}
\caption{Joint distribution in Scenario III.}\label{tab:joint3}
\begin{center}
\begin{tabular}{ m{4em}c|c|c|c} 
\multicolumn{2}{r}{\multirow{2}{*}{Scenario III. }} &\multicolumn{3}{c}{Evidence Suggests} \\
\multicolumn{2}{c}{} & (G) $\rand{x} = -1$ & (NG) $\rand{x} = 0$ & (I) $\rand{x} = 1$ \\ 
\cline{2-5}
\multirow{3}{*}{\shortstack{Prosecutor\\Thinks}} & (I) $\rand{y}=1$ & 0.05 & 0.05 & 0.1 \\ 
\cline{2-5}
& (NG) $\rand{y}=0$ & 0.05 & 0.15 & 0.1 \\ 
\cline{2-5}
& (G) $\rand{y}=-1$ & 0.2 & 0.2 & 0.1
\end{tabular}
\end{center}
\end{table}

{\cb Analogously, we can view prosecutor, defendant, and judge, respectively, as a sensor, a plant, and a controller (or an estimator) in a (non-cooperative) multi-agent system. Then the sensor would report the status of the plant strategically in order to induce the controller to perceive the status in a certain way (e.g., to deceive the controller) while the controller would only be interested in learning the plant's status.}

Note that in Scenario I, the prosecutor's intuition always says that the defendant is guilty,  which can also be viewed as the prosecutor always seeks for conviction similar to the example studied in \cite{ref:Kamenica11}. However, this differs from the example in \cite{ref:Kamenica11} in terms of the cost measures, and therefore also the result is different. The cost of \player{S} when \player{R} has the belief $\begin{bmatrix} 1-\mu & \mu \end{bmatrix}'$, where $\mu := \prob{\rand{z}=\begin{bmatrix} -1 & -1 \end{bmatrix}'}$, is given by
\begin{equation}
\hat{c}_S(\mu) = 4\mu^2 -8\mu + 3,
\end{equation} 
which is a convex function. Note that even though $\trace{\bar{V}\vec{p}\vec{p}'}$ is a non-convex function of $\vec{p}$, it turns out to be a convex function over the unit simplex under the specific settings of this example. However, as observed in the following examples, this is not always the case. Furthermore, as characterized in \cite{ref:Kamenica11}, since $\hat{c}_S(\cdot)$ is a convex function, null signaling is the optimal one in Scenario I. Alternatively, we have its convex envelope $C(\mu)=\hat{c}_S(\mu)$ and the minimum cost for \player{S} is given by $C(\mu) = \hat{c}_S(\mu_o) = 0.96$, where $\mu_o$ is the prior probability of $\begin{bmatrix} -1 & -1 \end{bmatrix}'$, i.e., $0.3$. 

On the other hand, in Scenarios II and III, the associated cost of \player{S} for a given posterior belief of \player{R} does not end up being a convex function. The dimensions of freedom to select the posterior are $3$ and $8$, respectively, in Scenarios II and III. Therefore, in order to apply the same solution concept, we need to compute the convex envelope of some non-convex functions defined over $\universe{R}^3$ and $\universe{R}^8$, instead of $\universe{R}$ as we do in Scenario I. However, through the proposed framework, we can efficiently compute the minimum cost for \player{S} and the associated optimal signaling strategies in those scenarios {\cb conclusively}. 

\begin{table}[t!]
\renewcommand{\arraystretch}{1.5}
\caption{\cb A table of $\E{\estimate{z}'V\estimate{z}}$  for different signaling strategies in Scenarios I-III within the context of deception}\label{tab:main}
\begin{center}
\begin{tabular}{ c|c|c|c}
Scenario (\# of States) & I (2) & II (4) & III (9) \\
\hline
\hline
Null Signaling  & {\bf 0.96} & 0.16 & 0 \\
\hline
Full Signaling  & 1.8 & 0.6 & 0.3 \\
\hline
{\cb LP-25 in \cite{ref:Dughmi16}} & {\cb 0.9615} & {\cb -0.4625} & {\cb -0.376} \\
{\cb (output)} & {\cb (0.8465)} & {\cb (-0.625)} & {\cb (-0.5299)}\\
\hline
{\cb LP-100 in \cite{ref:Dughmi16}} & {\cb\bf 0.96} & {\cb -0.4646} & {\cb -0.4128} \\
{\cb (output)} & {\cb (0.9319)} & {\cb (-0.5461)} & {\cb (-0.4933)} \\
\hline
Optimal Signaling  & {\bf 0.96} & {\bf -0.4715} & {\bf -0.4283} \\
\hline
\hline
Our SDP-relaxation & {\bf 0.96} & {\bf -0.4715} & \bf{-0.4283} \\
\hline
{\cb SDP-relaxation in \cite{ref:Tamura14}} & {\cb\bf 0.96} & {\cb -0.5411} & {\cb -0.4550}
\end{tabular}
\end{center}
\end{table}

In Table \ref{tab:main}, we tabulate the results.\footnote{In null, full, and optimal signaling, $\E{\estimate{z}'V\estimate{z}}$ is, respectively, equal to $\trace{\bar{V}\vec{p}_o\vec{p}_o'}$, $\trace{\bar{V}\diag{\vec{p}_o}}$, and $\trace{\bar{V}\Xi^*}$. {\cb Furthermore, we use the solvers provided by \cite{ref:CVX,ref:Grant08} at an ordinary personal computer in order to solve the associated LPs and SDPs.}} Proposition \ref{prop:interesting} says that the SDP-relaxation and the primal problem are equivalent for $n\leq 4$, as also seen in Table \ref{tab:main}. We also note that the SDP-relaxation has turned out to be a tight lower bound in Scenario III, where $n=9$. In Fig. \ref{fig:deception}, we illustrate the convergence behavior of the solutions of the inner and outer polyhedral approximations across the iterations. Note that with the increase of the size of the state space, the corresponding number of iterations necessary for convergence increase significantly while the computational complexity of each iteration increases further with an increase in the size of the state space.

{\cb In Table \ref{tab:main}, we also tabulate the results obtained via the existing methods that are explained in detail in Subsections \ref{sec:LP} and \ref{sec:Gauss}. For example LP-25 and LP-100 correspond to $\E{\estimate{z}'V\estimate{z}}$ attained for signaling strategies computed via \eqref{eq:LLP} when the receiver's action space is quantized uniformly into 25 and 100 bins, respectively, while the values in parentheses are the associated minimum attained in \eqref{eq:LLP}. As expected, finer granularity quantization, e.g., 100 bins in comparison to 25 bins, leads to smaller approximation error. Although solving the LP, described in \eqref{eq:LLP}, in the scenarios where the receiver has more than 100 actions pushes the limit of computational capabilities of an ordinary computer, even LP-100 could not compute the optimal signaling strategies with a negligible error when state space has sizes $4$ and $9$. Furthermore the outcomes of the SDP-relaxation in \cite{ref:Tamura14}, as described in \eqref{eq:cc}, are not tight in Scenarios II and III, contrary to our SDP-relaxation.}

\begin{table}[t!]
\renewcommand{\arraystretch}{1.5}
\caption{\cb A table of $\E{\estimate{z}'V\estimate{z}}$ for different signaling strategies in Scenarios I-III but now within the context of privacy}\label{tab:main_p}
\begin{center}
\begin{tabular}{ c|c|c|c}
Scenario (\# of States) & I (2) & II (4) & III (9) \\
\hline
\hline
Null Signaling & 0.84 & -0.16 & 0.09 \\
\hline
Full Signaling & {\bf 0} & 0 & 0.1 \\
\hline
{\cb LP-25 in \cite{ref:Dughmi16}} & {\cb\bf 0} & {\cb -0.9167} & {\cb -0.4469}\\
{\cb(output)} & {\cb\bf (0)} & {\cb (-0.975)} & {\cb (-0.4983)} \\
\hline
{\cb LP-100 in \cite{ref:Dughmi16}} & {\cb\bf 0} & {\cb -0.9432} & {\cb -0.4625} \\
{\cb(output)} & {\cb\bf (0)} & {\cb (-0.9531)} & {\cb (-0.4691)} \\
\hline
Optimal Signaling & {\bf 0} & {\bf -0.9583} & {\bf -0.4707} \\
\hline
\hline
Our SDP-relaxation & {\bf 0} & {\bf -0.9583} & {\bf -0.4707}\\
\hline
{\cb SDP-relaxation in \cite{ref:Tamura14}} & {\cb \bf 0} & {\cb -0.9780} & {\cb -0.4911}
\end{tabular}
\end{center}
\end{table}

We have also analyzed the cost of \player{S} {\cb within the context of privacy} over Scenarios I-III. We have tabulated the results in Table \ref{tab:main_p}. Similar to the previous case, we can apply the solution process introduced in \cite{ref:Kamenica11} to this problem for Scenario I. Correspondingly, the cost of \player{S} when \player{R} has the belief $\begin{bmatrix} 1-\mu & \mu \end{bmatrix}'$, where $\mu := \prob{\rand{z}=\begin{bmatrix} -1 & -1 \end{bmatrix}'}$, is now given by
\begin{equation}
\hat{c}_S(\mu) = - 4\mu^2+4\mu,
\end{equation} 
which is a concave function. As characterized in \cite{ref:Kamenica11}, since $\hat{c}_S(\cdot)$ is a concave function, we can conclude that full signaling is the optimal one in Scenario I. Alternatively, a geometrical inspection yields that the convex envelope of $\hat{c}_S(\mu)$ is given by $C(\mu)=0$, which coincides with the cost for the full signaling case as expected. {\cb We also note that in Scenario III, the output of LP-100 is larger than the optimal cost, which exemplifies that \eqref{eq:LLP} accompanied with a quantization of the receiver's action set does not bound the optimal cost from below.} 

\begin{figure}[t]
\centering
\includegraphics[width=.4\textwidth]{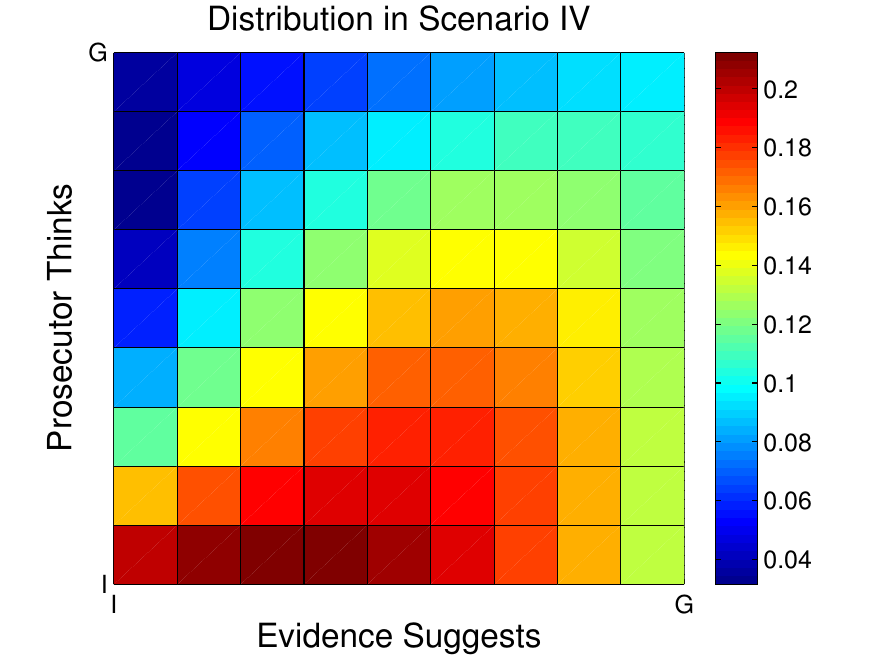}
\caption{\cb Joint distribution over the state space of size $100$ in which what evidence suggests and what the prosecutor thinks are scaled into $10$ uniform bins from ``Innocent" to ``Guilty"}\label{fig:scenario4}
\end{figure}

\begin{table}[t!]
\renewcommand{\arraystretch}{1.5}
\caption{\cb A table of $\E{\estimate{z}'V\estimate{z}}$ for full and null signaling in comparison to the lower bounds obtained via the SDP-relaxations within the context of deception}\label{tab:main_large}
\begin{center}
\cb
\begin{tabular}{ c|c}
Scenario (\# of States) & IV (100) \\
\hline
\hline
Null Signaling & 0.0146  \\
\hline
Full Signaling & 0.25 \\
\hline
LP-100 in \cite{ref:Dughmi16} (Output) & -0.2213 (-0.2822) \\
\hline
\hline
Our SDP-relaxation & -0.2365\\
\hline
SDP-relaxation in \cite{ref:Tamura14} & -0.2416
\end{tabular}
\end{center}
\end{table}

{\cb Finally we compare our SDP-relaxation with the SDP-relaxation in \cite{ref:Tamura14} and the LP formulation in \cite{ref:Dughmi16} for a 100-bin (uniform) quantization of the receiver's/judge's action space in Scenario IV where the size of the state space is $100$. For example, we set the state space and the joint distribution by interpolating the configuration in Scenario III. In Fig. \ref{fig:scenario4}, we plot the joint distribution over the state space. And in Table \ref{tab:main_large}, we tabulate the results. We note that $-0.2213$ is the value of $\E{\estimate{z}'V\estimate{z}}$ associated with the signaling strategy computed via LP-100. Therefore it can be viewed as an upper bound on the minimum cost. On the other hand, our SDP-relaxation is a lower bound on the minimum cost and again appears tight in comparison to the SDP-relaxation in \cite{ref:Tamura14}.}

\section{Concluding Remarks} \label{sec:conclusion}

{\cb We have addressed the Bayesian persuasion problem with state-dependent quadratic cost measures for arbitrary distributions with finite support. This problem setting has applications, e.g., varying from deception to privacy, in multi-agent (non-cooperative) systems. However, it is challenging to compute optimal signaling strategies or optimal signaling cost via existing solution concepts effectively even for distributions that have support over a small state space, since the associated optimization problem turns out to be highly nonlinear and non-convex. To mitigate this issue, we have formulated an equivalent linear optimization problem over the cone of completely positive matrices by exploiting the quadratic structure of the cost measures and showed that its (strong) dual is a linear optimization problem over the cone of copositive matrices. Then we have exemplified how one can adopt the existing solution methods on copositive programming to solve the optimization problem effectively, and we have examined its performance over various scenarios numerically. We have also quantified the approximation error for a quantized version of a continuous distribution and provided an SDP relaxation of the optimization problem, which can be solved for relatively larger state spaces and therefore can be used as a benchmark lower bound for the minimum signaling cost in signaling problems with large (or continuous) state spaces. 

Some of the future research directions include development of computational tools specific to the configuration of the equivalent problem, characterization of the hardness of the signaling problem, and its generalization to dynamic and/or noisy settings with multiple senders and/or multiple receivers.}

\section*{Acknowledgment}
We thank the anonymous reviewers and the associated editor for the insight they have provided, and for their constructive comments.

\appendices

\section{Proof of Proposition \ref{prop:main}} \label{app:main}
\begin{claim}\label{claim:if}
For any signaling rule $\Srule\in\Sspace$, $\myXi\in\universe{S}^n$ satisfies
\begin{itemize}
\item $\myXi\in\CP^n$, 
\item $\myXi\vec{1} = \vec{p}_o$,
\end{itemize}
where $\vec{p}_o := \begin{bmatrix} \prior(z_1) &\ldots&\prior(z_n)\end{bmatrix}'$.
\end{claim}

\begin{myproof}
We can decompose $\myXi\in\universe{S}^n$, as described in \eqref{eq:entry}, as $\myXi = AA'$, where
\begin{equation}\label{eq:A}
A := \begin{bmatrix} p_{\signal_1}(z_1)p(\signal_1)^{1/2} & \cdots & p_{\signal_k}(z_1)p(\signal_k)^{1/2} \\ \vdots & & \vdots \\ p_{\signal_1}(z_n)p(\signal_1)^{1/2} & \cdots & p_{\signal_k}(z_n)p(\signal_k)^{1/2} \end{bmatrix},
\end{equation}
which is clearly a nonnegative matrix since all the entries are products of (nonnegative) probability measures. This yields that $\myXi\in\CP^n$.

For a given signaling rule $\Srule\in\Sspace$, let $\signalsSet_o\subseteq\signalsSet$ denote the set of signals that \player{S} sends with positive probability, i.e., $p(\signal)>0$ for all $\signal\in\signalsSet_o$. Then, the sum of entries of $\myXi$ at the $i$th row is given by 
\begin{align}
\sum_{j=1}^n \myXi[i,j] &\stackrel{(a)}{=} \sum_{\signal\in\signalsSet_o} p(\signal)\posterior(z_i) \overbrace{\sum_{j=1}^n\posterior(z_j)}^{=1},\\
&\stackrel{(b)}{=} \sum_{\signal\in\signalsSet_o} \cancel{p(\signal)}\frac{\Srule(\signal|z_i)\prior(z_i)}{\cancel{p(\signal)}},\\
&= \prior(z_i) \sum_{\signal\in\signalsSet_o} \Srule(\signal|z_i),\\
&\stackrel{(c)}{=} \prior(z_i),\label{eq:equalPrior}
\end{align}
where $(a)$ follows from \eqref{eq:posterior} and \eqref{eq:signalProb}, $(b)$ follows from \eqref{eq:posterior}, and $(c)$ follows since $\Srule(\cdot|z)\in\Delta(\signalsSet)$. By \eqref{eq:equalPrior}, we have $\myXi\vec{1} = \vec{p}_o$, which completes the proof of the claim.
\end{myproof}

\begin{claim}\label{claim:onlyIf}
For any completely positive matrix $\Xi\in\CP^n$ that satisfies $\Xi\vec{1} = \vec{p}_o$, there exists a signaling rule $\Srule\in\Sspace$ such that $\myXi = \Xi$, where $\myXi\in\CP^n$ is as described in \eqref{eq:entry}.
\end{claim}

\begin{myproof}
Consider any completely positive matrix $\Xi\in\CP^n$ that satisfies $\Xi\vec{1} = \vec{p}_o$. By the definition of completely positive matrices, we can decompose $\Xi\in\CP^n$ into $\Xi = BB'$, where 
\begin{equation}
B=\begin{bmatrix} \vec{b}_1&\ldots&\vec{b}_k \end{bmatrix}\in\universe{R}_+^{n\times k}
\end{equation} 
is some nonnegative matrix. We note that the decomposition is not necessarily a unique one \cite{ref:Berman03}. For example, by padding zero columns into $B$, we can generate infinitely many decompositions. Correspondingly, we assume, without loss of generality, that in the decomposition of $\Xi\in\CP^n$, the nonnegative matrix $B\in\universe{R}_+^{n\times k}$ does not have an all zero column, i.e., $\vec{b}_i\neq \vec{0}$ for all $i=1,\ldots,k$.

Recall that for any given signaling rule $\Srule\in\Sspace$, we can decompose $\myXi = AA'$, where the matrix $A\in\universe{R}_+^{n\times |\signalsSet|}$ is as described in \eqref{eq:A}. Correspondingly, if we can show that there exists a signaling rule $\Srule\in\Sspace$ such that $A = B$, then this would imply that $\Xi = \myXi$ for that signaling rule. To this end, we let $|\signalsSet|=k$, and introduce auxiliary vectors 
\begin{equation}
\vec{\Srule}_i := \begin{bmatrix} \Srule(\signal_i|z_1) & \ldots & \Srule(\signal_i|z_n)\end{bmatrix}',
\end{equation}
which, by \eqref{eq:signalProb}, yields that $p(\signal_i) = \vec{\Srule}_i' \vec{p}_o$. Therefore, by substituting \eqref{eq:posterior} in \eqref{eq:A}, we can write the matrix $A$ as
\begin{equation}
A = P_o \begin{bmatrix} \frac{\vec{\Srule}_1}{\sqrt{\vec{\Srule}_1'\vec{p}_o}} & \ldots & \frac{\vec{\Srule}_k}{\sqrt{\vec{\Srule}_k'\vec{p}_o}}\end{bmatrix},
\end{equation}
where $P_o := \mathrm{diag}\{\vec{p}_o\}$, which is nonsingular since the prior distribution $\prior$ has complete support on $\stateSet$. Correspondingly, $A$ would be equal to $B$ provided that
\begin{equation}\label{eq:equality}
\frac{P_o\vec{\Srule}_i}{\sqrt{\vec{\Srule}_i'\vec{p}_o}} = b_i,\;\forall i = 1,\ldots,k,
\end{equation}
and it can be verified that we have \eqref{eq:equality} if
\begin{equation}\label{eq:candidate}
\vec{\Srule}_i = \vec{b}_i'\vec{1} P_o^{-1} \vec{b}_i.
\end{equation}

However, we also need to inspect the validity of \eqref{eq:candidate} as a signaling rule, i.e., whether $\Srule(\cdot|z)\in\Delta(\signalsSet)$ for each $z\in\stateSet$ or not. To this end, we focus on the equivalent conditions that $\Srule(\signal|z)$ for all $\signal\in\signalsSet$ and $z\in\stateSet$, and $\sum_{\signal\in\signalsSet}\Srule(\signal|z) = 1$ for each $z\in\stateSet$. The former condition is satisfied by the definition, since $\vec{b}_i$ is a column of the nonnegative matrix $B\in\universe{R}_+^{n\times k}$. Verification of the latter condition is relatively more involved. To this end, let us introduce
\begin{equation}
\myPi := \begin{bmatrix} \vec{\Srule}_1 & \ldots & \vec{\Srule}_k \end{bmatrix}.
\end{equation}
Then, the latter constraint is equivalent to $\myPi\vec{1} = \vec{1}$, i.e., the sum of columns of $\myPi$ is an all $1$s vector. If we set $\vec{\Srule}_i$ as in \eqref{eq:candidate}, then we would obtain
\begin{equation}
\myPi = \begin{bmatrix} (\vec{b}_1'\vec{1}) P_o^{-1} \vec{b}_1 & \ldots & (\vec{b}_k'\vec{1}) P_o^{-1} \vec{b}_n\end{bmatrix},
\end{equation}
and correspondingly the sum of columns of $\myPi$ is given by
\begin{align}\label{eq:step1}
\myPi\vec{1} = P_o^{-1} \sum_{i=1}^k \vec{b}_i (\vec{b}_i'\vec{1}).
\end{align}
Note that $(\vec{b}_i'\vec{1})P_o^{-1}\vec{b}_i = P_o^{-1}\vec{b}_i(\vec{b}_i'\vec{1})$ since $\vec{b}_i'\vec{1}$ is just a scalar. 

Recall that the completely positive matrix $\Xi\in\CP^n$ satisfies $\Xi\vec{1} = \vec{p}_o$, which can also be written as
\begin{equation}\label{eq:step2}
BB'\vec{1} = \vec{p}_o \Leftrightarrow \sum_{i=1}^k \vec{b}_i\vec{b}_i'\vec{1} = \vec{p}_o.
\end{equation}
Therefore, by \eqref{eq:step1} and \eqref{eq:step2}, we obtain $\myPi\vec{1} = P_o^{-1}\vec{p}_o = \vec{1}$, and correspondingly \eqref{eq:candidate} is a valid signaling strategy, which completes the proof of the claim.
\end{myproof}

Based on \eqref{eq:XiEquivalent}, Claims \ref{claim:if} and \ref{claim:onlyIf} yield \eqref{eq:CPequivalent}, and \eqref{eq:optimalSignal} follows from \eqref{eq:candidate}, which completes the proof.

\section{Proof of Proposition \ref{prop:duality}}\label{app:duality}

{\cb In the following we first formulate the dual problem, and then show the strong duality.}

{\cb {\bf Part $i)$ Dual Problem.} As pointed out in \cite[Chapter 4.7]{ref:Gartner12}, given a primal cone program:
\begin{align} \label{eq:primalw}
\max_{x} \;\langle c,x\rangle \mbox{ s.t. } b-A(x) \in L\mbox{ and }x\in K,
\end{align}
where $K,L$ are cones and $A(\cdot)$ is a linear operator, its dual program is given by
\begin{align}\label{eq:dualw}
\min_{y}\; \langle b,y\rangle \mbox{ s.t. } &A^T(y) -c \in K^*\mbox{ and }
y\in L^*,
\end{align}
where $A^T(\cdot)$ is the adjoint of the linear operator $A(\cdot)$.
Based on the fact that $\min_x f(x) = \max_x - f(x)$, our primal problem can be written as
\begin{equation}\label{eq:probw}
\max_{\Xi} \;\trace{\Xi (-\bar{V})}\mbox{ s.t. } \vec{p}_o - \Xi\vec{1} = \vec{0}\mbox{ and } \Xi\in\CP^n.
\end{equation}
Within the framework of \eqref{eq:primal}, let $x=\Xi$, $c = -\bar{V}$, $\langle c,x\rangle = \trace{c^Tx}$, $A(\Xi) = \Xi\vec{1}$, $b = \vec{p}_o$, $L = \{\vec{0}\}$, and $K=\CP^n$. Note that 
\begin{equation}
\langle y,A(\Xi)\rangle = \trace{y^T\Xi\vec{1}} = \trace{\vec{1}y^T \Xi},
\end{equation}
which yields that $A^T(y) = \vec{1}y^T$ is an adjoint of the linear operator $A(\Xi) = \Xi\vec{1}$. Furthermore, the dual of the closed convex cones $L=\{\vec{0}\}\subset\mathbb{R}^n$ and $K=\CP^n\subset\mathbb{S}^n$ with respect to the trace inner product are, respectively, given by 
\begin{align}
&L^* = \left\{y\in \mathbb{R}^n: \trace{y^T x} \geq 0,\forall x\in L\right\} = \mathbb{R}^n\\
&K^* = \left\{y\in \mathbb{S}^n: \trace{y^T x} \geq 0,\forall x\in K\right\} = \COP^n.
\end{align}
Therefore, \eqref{eq:dualw} yields that the dual of \eqref{eq:probw} is given by
\begin{equation}
\min_{y}\; \vec{p}_o^Ty \mbox{ s.t. } \vec{1}y^T+\bar{V} \in \COP^n \mbox{ and } y\in \mathbb{R}^n,
\end{equation}
or equivalently if we set $\tilde{y} = - y$, we obtain
\begin{equation*}
\max_{\tilde{y},S}\; \vec{p}_o^T\tilde{y} \mbox{ s.t. } -\vec{1}\tilde{y}^T+\bar{V} = S,\, \tilde{y}\in \mathbb{R}^n,\mbox{ and } S\in\COP^n.
\end{equation*}}

{\cb \bf Part $ii)$ Strong Duality.} The primal problem is feasible since the constraint set is not empty based on Claims \ref{claim:if} and \ref{claim:onlyIf} {\cb in Appendix \ref{app:main}}.

The fact that the primal problem has finite value follows by the extreme value theorem since the optimization objective is linear in the optimization argument and the constraint set 
\begin{equation}
\{\Xi\in\CP^n | \Xi\vec{1} = \vec{p}_o\}
\end{equation} 
is a closed and bounded subset of $\CP^n$.

It is relatively more involved to show that the primal problem entails an interior point. Particularly, a characterization of the interior of $\CP^n$ is given by \cite[Theorem 3.3]{ref:Dickinson11}
\begin{align}
\interior{\CP^n} = \{AA' \,|\, &\rank{A}=n,A=\begin{bmatrix} \vec{a} & \tilde{A}\end{bmatrix}\nn\\
&\ni \vec{a}>0, \tilde{A}\geq 0\}.
\end{align}
Therefore, the question is whether there exists a $\Xi\in\interior{\CP^n}$ such that $\Xi\vec{1}=\vec{p}_o$. 

Based on Claims \ref{claim:if} and \ref{claim:onlyIf}, let us consider the associated signaling problem where we set the signal space as $\signalsSet = \{\signal_o = \varnothing,\signal_1=z_1,\ldots,\signal_n=z_n\}$, and the prior distribution over $\stateSet$ to have full support, without loss of generality. Consider the two extreme cases: full disclosure and null disclosure, respectively, given and denoted by $\overline{\Srule}(z_i) = z_i$, and $\underline{\Srule}(z_i) = \varnothing$ for all $i=1,\ldots,n$. Note that for a given signaling strategy $\Srule\in\Sspace$, an entry of the associated completely positive matrix $\myXi$ is described in \eqref{eq:entry}. Furthermore, a component of a decomposition of $\myXi$ is described in \eqref{eq:A}. Correspondingly, we obtain
\begin{align}
&\Xi_{\overline{\Srule}} = \overline{A}\overline{A}' \ni \overline{A} := \begin{bmatrix} \vec{0} & P_o \end{bmatrix},\\
&\Xi_{\underline{\Srule}} = \underline{A}\underline{A}' \ni \underline{A} := \begin{bmatrix} \vec{p}_o & O \end{bmatrix}.
\end{align}
This yields that if \player{S} selects a signaling strategy $\Srule\in\Sspace$ that discloses $z\in\stateSet$ truthfully, i.e., $\signal_i=z_i$, with probability $\lambda\in(0,1)$ and discloses $\signal_o$ otherwise, then we obtain
\begin{equation}
\myXi = AA' \ni A := \begin{bmatrix} (1-\lambda)\vec{p}_o & \lambda P_o\end{bmatrix},
\end{equation}
in which the first column is a positive vector and $\rank{A}=n$ since $p_o$ is an all-positive vector, i.e., the prior distribution has full support over $\stateSet$ by the formulation. This yields that $\myXi\in\interior{\CP^n}$ and $\myXi\vec{1}=\vec{p}_o$.

Since the conditions for the strong duality theorem \cite[Theorem 4.7.1]{ref:Gartner12} hold, we have strong duality between the primal and dual problems, which concludes the proof.

\section{Proof of Corollary \ref{corollary:quantization}}\label{app:quantization}

We first note that the following inequality always holds
\begin{align}\label{eq:always}
\min_{\rand{s}} \trace{\E{\estimate{z}\estimate{z}'}V} \leq \min_{\rand{s}} \trace{\E{\estimate{z}_q\estimate{z}_q'}V}
\end{align}
since any quantization would restrict \player{S}'s strategy space for continuous distributions.

Let $\estimate{e} = \estimate{z}-\estimate{z}_q$ for a given signal $\rand{\signal}$, i.e., $\estimate{e} = \E{\rand{e}|\rand{\signal}}$. Then, for any signal $\rand{\signal}\sim\stateSet$, we have
\begin{align}
\E{\estimate{z}\estimate{z}'} = \E{\estimate{z}_q\estimate{z}_q'}+ \E{\estimate{z}_q\estimate{e}'} + \E{\estimate{e}\estimate{z}_q'} + \E{\estimate{e}\estimate{e}'}.\nn
\end{align}
Correspondingly, we obtain
\begin{align}
&\min_{\rand{s}} \trace{\E{\estimate{z}\estimate{z}'}V} \geq \min_{\rand{s}} \trace{\E{\estimate{z}_q\estimate{z}_q'}V} \nn\\
&\hspace{.15in}+ \min_{\rand{s}} \trace{(\E{\estimate{z}_q\estimate{e}'} + \E{\estimate{e}\estimate{z}_q'} + \E{\estimate{e}\estimate{e}'})V}.
\end{align}
Let us take a closer look at the second term on the right-hand-side, which can also be written as
\begin{align}\label{eq:last}
-\max_{\rand{s}} -\trace{(\E{\estimate{z}_q\estimate{e}'} + \E{\estimate{e}\estimate{z}_q'} + \E{\estimate{e}\estimate{e}'})V}
\end{align}
Then, the Cauchy-Schwarz inequality for random vectors yields that \eqref{eq:last} is bounded from above by
\begin{equation}
-2\E{\estimate{z}_q'V\estimate{e}} - \E{\estimate{e}'V\estimate{e}} \leq (2\|\estimate{z}_q\| + \|\estimate{e}\|) \|V\|_2\|\estimate{e}\|,\nn
\end{equation}
since $\|-V\|_2 = \|V\|_2$. Note that the right-hand-side depends on the signal $\rand{s}$. However, we also have
\begin{align}
&\|\rand{z}_q-\estimate{z}_q\|^2 =\|\rand{z}_q\|^2 - \|\estimate{z}_q\|^2\geq 0,\label{eq:zqbound}\\
&\|\rand{e}-\estimate{e}\|^2 =\|\rand{e}\|^2 - \|\estimate{e}\|^2\geq 0.\label{eq:ebound}
\end{align}
Therefore, by \eqref{eq:zqbound} and \eqref{eq:ebound}, we obtain
\begin{equation}\label{eq:upp}
-2\E{\estimate{z}_q'V\estimate{e}} - \E{\estimate{e}'V\estimate{e}} \leq (2\|\rand{z}_q\| + \|\rand{e}\|) \|V\|_2\|\rand{e}\|,
\end{equation}
where, now, the right-hand-side does not depend on the signal. Therefore, we obtain
\begin{align}\nn
\epsilon \geq \min_{\rand{s}} \trace{\E{\estimate{z}_q\estimate{z}_q'}V}-\min_{\rand{s}} \trace{\E{\estimate{z}\estimate{z}'}V}\geq 0,
\end{align}
where $\epsilon$ is as described in \eqref{eq:epsilon}, which completes the proof.

\bibliographystyle{IEEEtran}
\bibliography{ref}

\begin{IEEEbiographynophoto}{Muhammed O. Sayin}  is currently a postdoctoral associate in Laboratory for Information and Decision Systems in Massachusetts Institute of Technology. He received the Ph.D. degree in Electrical and Computer Engineering from the University of Illinois at Urbana-Champaign (UIUC) in 2019. He received the B.S. and M.S. degrees in Electrical and Electronics Engineering from Bilkent University, Ankara, Turkey, in 2013 and 2015, respectively. His current research interests include signaling games, dynamic games and decision theory, learning in game theory and multi-agent systems.
\end{IEEEbiographynophoto}

\begin{IEEEbiographynophoto}{Tamer Ba\c{s}ar} (S'71-M'73-SM'79-F'83-LF'13) is with the University of Illinois at Urbana-Champaign, where he holds the academic positions of  Swanlund Endowed Chair; Center for Advanced Study Professor of  Electrical and Computer Engineering; Research Professor at the Coordinated Science Laboratory; and Research Professor  at the Information Trust Institute. He is also the Director of the Center for Advanced Study.

He received B.S.E.E. from Robert College, Istanbul, and M.S., M.Phil, and Ph.D. from Yale University. He is a member of the US National Academy of Engineering,  member of the  European Academy of Sciences, and Fellow of IEEE, IFAC (International Federation of Automatic Control) and SIAM (Society for Industrial and Applied Mathematics), and has served as president of IEEE CSS (Control Systems  Society), ISDG (International Society of Dynamic Games), and AACC (American Automatic Control Council). He has received several awards and recognitions over the years, including the highest awards of IEEE CSS, IFAC, AACC, and ISDG, the IEEE Control Systems Award, and a number of international honorary doctorates and professorships. He has over 900 publications in systems, control, communications, and dynamic games, including books on non-cooperative dynamic game theory, robust control, network security, wireless and communication networks, and stochastic networked control. He was the Editor-in-Chief of Automatica between 2004 and 2014, and is currently  editor of several book series. His current research interests include stochastic teams, games, and networks; distributed algorithms; security; and cyber-physical systems.
\end{IEEEbiographynophoto}
\vfill

\end{document}